\documentclass{moriond}

\def\be{\begin{equation}}
\def\ee{\end{equation}}
\def\bea{\begin{eqnarray}}
\def\eea{\end{eqnarray}}

\def\iso#1#2{\mbox{${}^{#2}{\rm #1}$}}
\def\he#1{\iso{He}{#1}}
\def\li#1{\iso{Li}{#1}}
\def\dpg{$d(p,\gamma)$\he3~}
\def\nnu{N_\nu}
\def\Ombh2{\Omega_{\rm b} h^2}

\def\dpg{$d(p,\gamma)$\he3~}

\newcommand{\Photo}{}

\begin{document}
\begin{flushright}
{\tt UMN-TH-4013/21, FTPI-MINN-21/06} \\
\end{flushright}

\vspace*{3.8cm}
\title{Impact of Current Results on Nucleosynthesis$^*$}

\author{ Keith A. Olive }

\address{William I. Fine Theoretical Physics Institute, School of
 Physics and Astronomy, \\  University of Minnesota, Minneapolis,
 Minnesota 55455, USA \\
  $^*$To be published in the proceedings of the 2021 EW session of the 55th Rencontres de Moriond.}

\maketitle

\abstracts{
The impact of recent results on Big Bang Nucleosynthesis is assessed. These
include the Planck likelihood distributions for the baryon density; recent progress in helium abundance determinations; and a recent cross section measurement for \dpg.}

\section{Introduction}

As one of the deepest probes of early universe cosmology, Big Bang Nucleosynthesis (BBN) \cite{bbn,cfo1,coc,cyburt,iocco,CFOY,coc18,FOYY},
has the ability to set constraints on a wide spectrum of extensions of the Standard Model \cite{kk,lisi,cfos,ms}. These can
often be cast in the form of a constraint on the number of neutrino flavors exceeding the Standard Model value of 3. 
These limits only have meaning if we are confident that BBN accurately accounts for the abundances of the light elements.
This concordance in turn depends on accurate measurements of the baryon density, and cross section measurements,
and of course accurate astrophysical abundance determinations.

The baryon density can be accurately determined from cosmic microwave background (CMB) anisotropy measurements. 
Beginning with WMAP \cite{wmap1} and more recently Planck \cite{Planck2018}, within the standard model, BBN is effectively a 
parameter free theory \cite{cfo2}.  Precise knowledge of the baryon density allows for well defined likelihood distributions for the BBN
predictions \cite{CFOY,FOYY,kk,kr,fo,cfo3}. These in turn can be convolved with observational likelihood functions to set constraints on the number of neutrino flavors. 

In this note, I will highlight the impact of 1) recent Planck data, 2) recent progress in \he4 abundance determinations, and 3) recent cross section measurements of \dpg.

\section{The Impact of the Planck Likelihoods}

\begin{figure}[!htb]
\begin{center}
\includegraphics[width=0.40\textwidth]{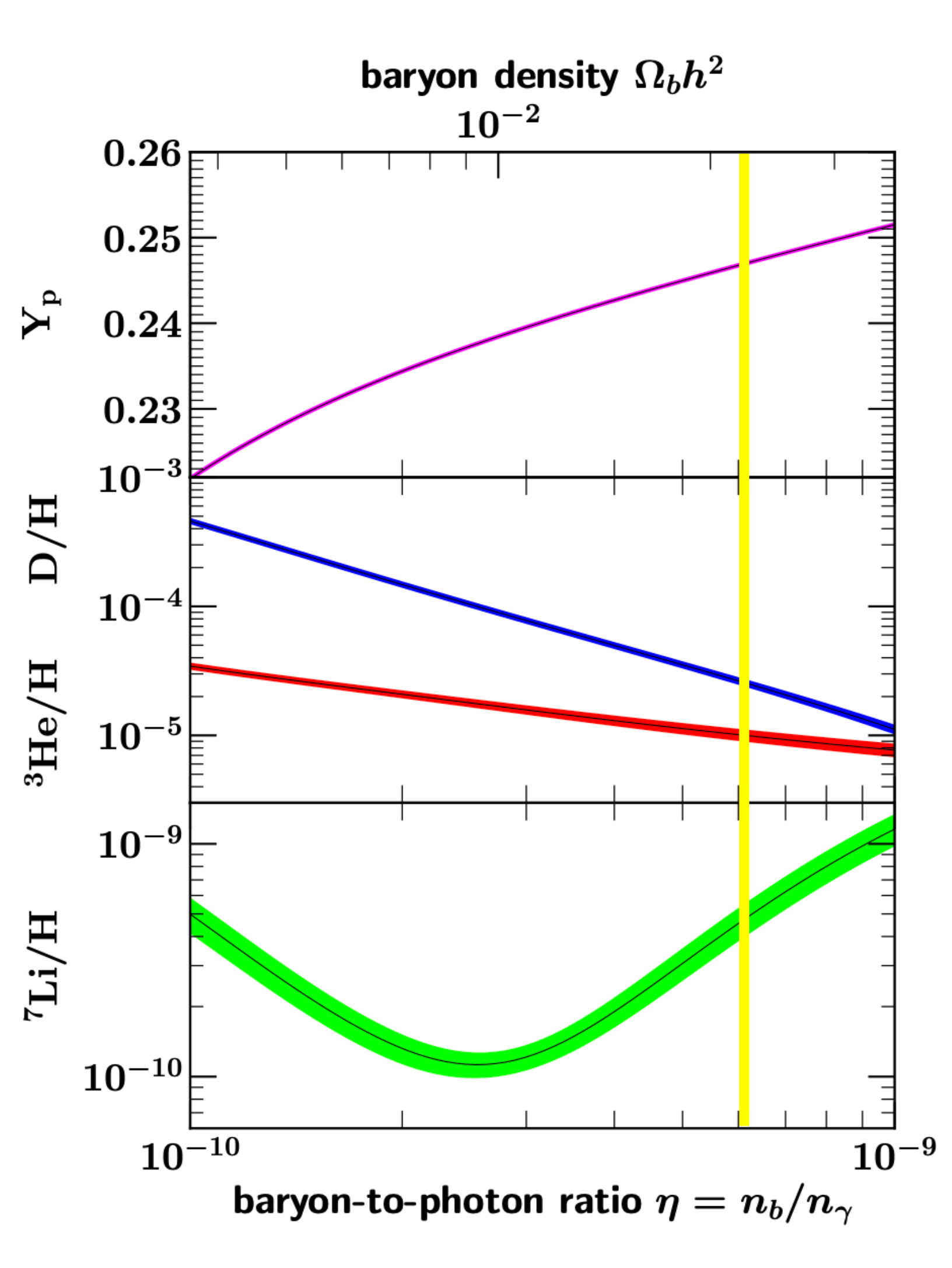}
\hskip .1in
\includegraphics[width=0.40\textwidth]{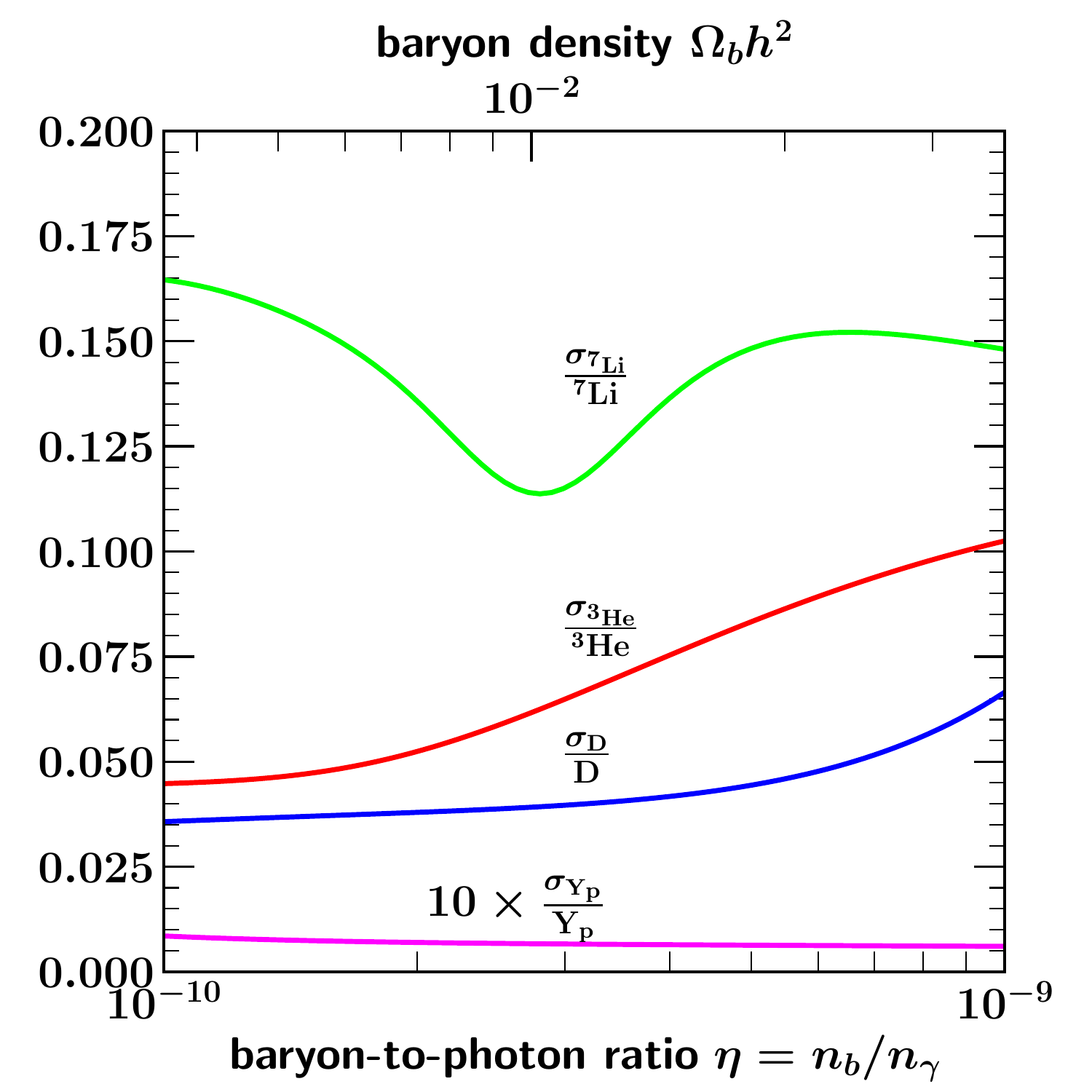}
\caption{
Left: Primordial abundances of the light nuclides as a function of
cosmic baryon content, as predicted by  standard BBN (``Schramm plot'').
Curve widths show $1\sigma$ errors.
Right: Fractional uncertainties in the light element abundance predictions shown on the left.  For each species $i$, we plot ratio of the standard deviation $\sigma_i$ to the mean $\mu_i$, as a function of baryon-to-photon ratio. The relative uncertainty of the \he4 abundance has been multiplied by a factor of 10.
\label{fig:schramm}
}
\end{center}
\end{figure}

Certainly the most revolutionary change in BBN analyses has been the ability to accurately fix the baryon density from CMB observations.
Standard model results for BBN are shown in Fig.~\ref{fig:schramm} which plots the light element abundances
as a function of the baryon density, $\Ombh2$ (upper scale) and $\eta = n_b/n_\gamma$ (lower scale).
In the left panel,
the mass fraction, $Y$, for He is shown while the abundances of 
D, \he3, and \li7 by number are shown relative to H. 
The line through each colored band is the mean value as a function of $\eta$ \cite{FOYY}. 
The thickness of each curve depicts the $\pm 1 \sigma$ spread in the predicted abundances.  The relative uncertainty,
(the thickness of the curves, relative to the central value) is shown more clearly in the right panel.

The vertical line in Fig.~\ref{fig:schramm}, shows the Planck determined value of the baryon density $\eta = (6.104 \pm 0.055) \times 10^{-10})$.
As one can see, the relative thinness of the Planck line implies, that there are definite predictions from BBN for each of the 
light element abundances. 

\begin{figure}[!htb]
\begin{center}
\includegraphics[width=0.40\textwidth]{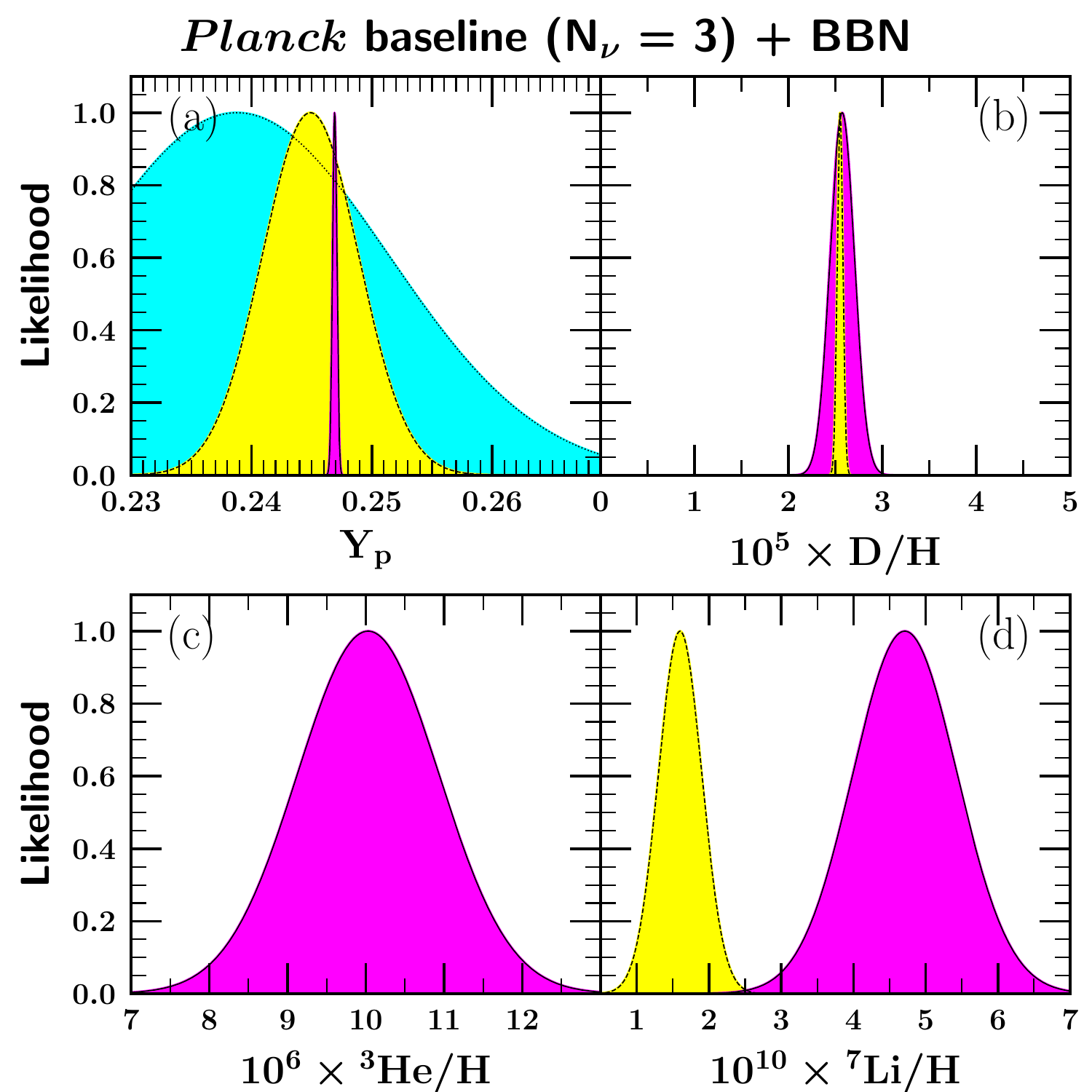}
\hskip .1in
\includegraphics[width=0.40\textwidth]{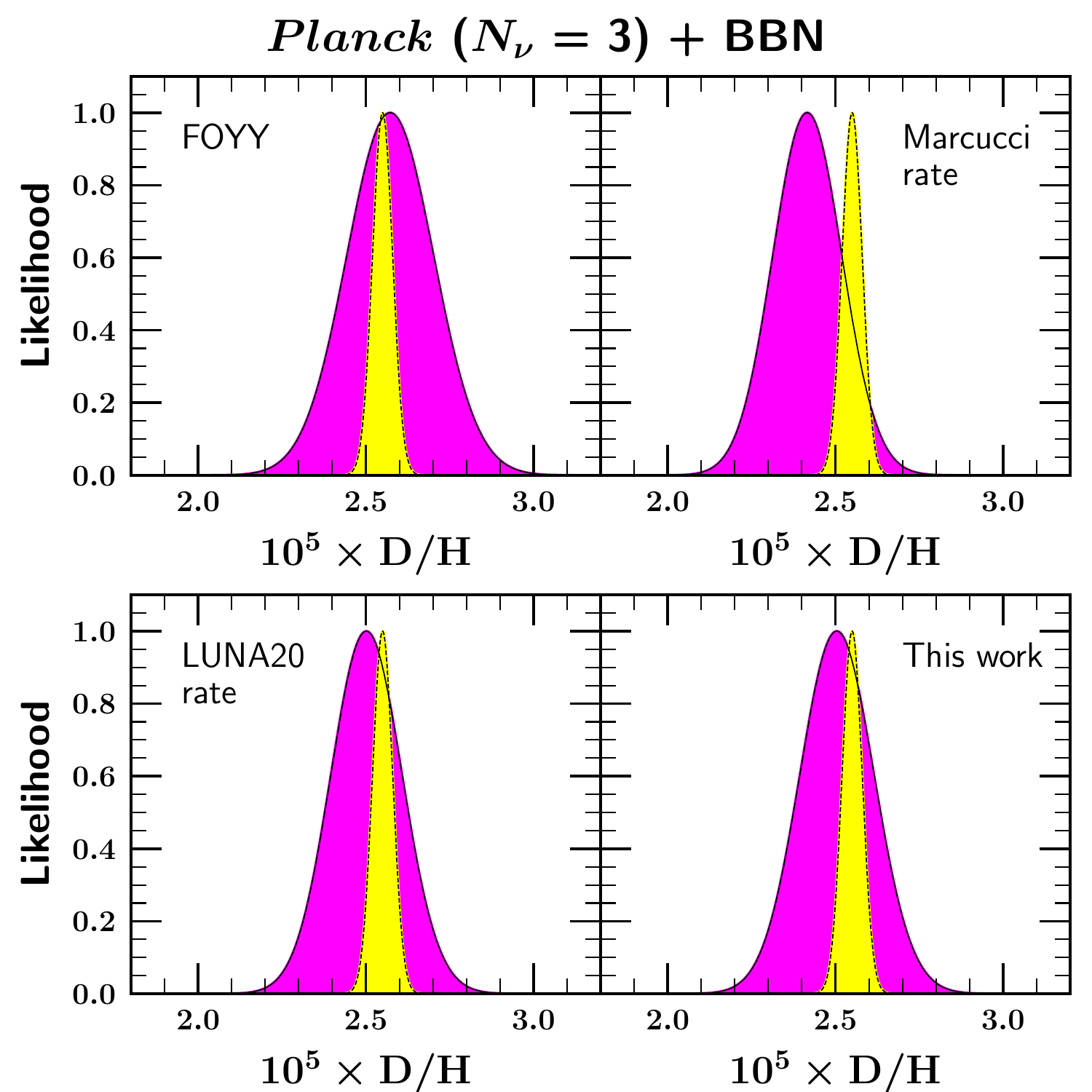}
\caption{
Light element abundance likelihood functions.  Shown on the left are likelihoods for each of the light nuclides,
normalized to a maximum value of 1.  The
dark-shaded (purple) curves are 
the BBN+CMB predictions, based on {\em Planck} inputs as discussed
in the text.  The light-shaded 
(yellow) curves show astronomical measurements
of the primordial abundances.  For 
\he4, the medium-shaded (cyan) curve shows
the independent CMB determination of \he4. In the right panel, the effect of the \dpg cross section is shown as described in the text.
\label{fig:2x2abs_2d}
}
\end{center}
\end{figure}

More formal analysis requires the construction of likelihood functions. 
Planck has provided many likelihood chains from CMB anisotropy and other measurements. 
We have taken the Planck likelihood chains
which are based on temperature and polarization data, T+TE+EE+lowE,
as well as CMB lensing  \cite{FOYY}. Furthermore, we use the chains that do not assume any BBN relation between the helium abundance and baryon density.
Thus for fixed $\nnu$, we have a likelihood function ${\mathcal L}_{\rm CMB}(\eta,Y_p)$ (and a separate likelihood function,  ${\mathcal L}_{\rm NCMB}(\eta, N_\nu, Y_p)$,
when $\nnu$ is not fixed). In addition, from a Monte Carlo, we can construct a BBN likelihood function for each element, $X$, 
${\mathcal L}_{\rm BBN}(\eta;X)$ (and similarly ${\mathcal L}_{\rm NBBN}(\eta, N_\nu;X)$ when $N_\nu$ is not fixed), and from the observations, ${\mathcal L}_{\rm OBS}(X)$. BBN results are obtained by
a convolution of the CMB and BBN likelihoods by integrating over $\eta$. The result for each light element is shown in Fig.~\ref{fig:2x2abs_2d} (left)
by the purple shaded curves. The yellow-shaded curves correspond to the observational likelihood with the exception of \he3 where 
reliable primordial abundance measurements do not exist. Interestingly, by integrating ${\mathcal L}_{\rm CMB}$ alone over $\eta$, 
we can determine the CMB likelihood for $Y_p$. This is shown by the cyan shaded curve. 
We see excellent agreement for D/H, good agreement for \he4, and strong discrepancy in \li7 constitutes the persistent lithium problem \cite{cfo5}.  

While the CMB plays a dominant role in determining the baryon density, BBN is able to refine this determination.
In Table~\ref{tab:eta}, the mean values of $\eta_{10} = 10^{10} \eta$ from the various convolutions of likelihood functions are given with their
1 $\sigma$ uncertainties.  Also shown is the value of $\eta_{10}$ at the peak of the likelihood.  As one can infer from the BBN results shown in 
Fig.~\ref{fig:schramm}, BBN and the observed helium abundance alone is not very apt at fixing $\eta$. Deuterium does much better, though the uncertainty in $\eta$ is still 4 times greater than the CMB alone. The primary cause for a deviation in the CMB value of $\eta$ is including
the BBN relation between $Y_P$ and $\eta$. This case is labeled CMB+BBN, and does not use any abundance measurements. 
The last line of the table gives the final result which includes the abundance measurements of D and He, with $\eta_{10} = 6.129 \pm 0.040$. 
One can repeat the process when $N_\nu$ is not held fixed \cite{FOYY} and the best fit value of $\eta_{10}$ drops to $6.090 \pm 0.055$. By marginalizing over
$\eta$, we can obtain likelihood functions for $N_\nu$, giving a mean value of $2.843 \pm 0.154$ when all
likelihoods are convolved. 

\begin{table}[!htb]
\caption{The mean value (and its uncertainty) of the baryon-to-photon ratio as well as the value of $\eta$ at the peak of the distribution using different combinations of observational constraints.
\label{tab:eta}
}
\begin{center}
\begin{tabular}{|l|c|c|}
\hline
 Constraints Used & mean $\eta_{10}$ & peak $\eta_{10}$ \\
\hline
CMB-only & $6.104\pm 0.055$ & 6.104 \\
\hline
\hline
BBN+$Y_p$ & $6.741 {}^{+1.220}_{-3.524}$ & 4.920 \\
\hline
BBN+D & $6.148\pm 0.191$ & 6.145 \\
\hline
BBN+$Y_p$+D & $6.143\pm 0.190$ & 6.140 \\
\hline
CMB+BBN & $6.129\pm 0.041$ & 6.129 \\
\hline
CMB+BBN+$Y_p$ & $6.128\pm 0.041$ & 6.128 \\
\hline
CMB+BBN+D & $6.130\pm 0.040$ & 6.129 \\
\hline
\hline
CMB+BBN+$Y_p$+D & $6.129\pm 0.040$ & 6.129 \\
\hline
\end{tabular}
\end{center}
\end{table}

\section{Towards Precision \he4 abundance determinations}

The \he4 abundance is determined from emission line data from extragalactic HII regions \cite{its07}. 
To account for systematics, a Markov-Chain Monte Carlo approach using the \he4 abundance as one of the several input
parameters has been adopted \cite{AOS2,AOS3}.  The \he4 abundance is determined simultaneously with several other physical input parameters which include the electron density, temperature, optical depth, neutral hydrogen fraction, as well as parameters for underlying absorption and reddening. The measured flux of six He and  three H lines (relative to H$\beta$), 
are compared with model predictions to obtain a $\chi^2$ statistic. The importance
of an additional IR line for \he4 was proposed \cite{itg14} and when available showed marked improvements in the accuracy of the \he4 abundance determinations \cite{AOS4}.

\begin{figure}[!htb]
\begin{center}
\includegraphics[width=0.40\textwidth]{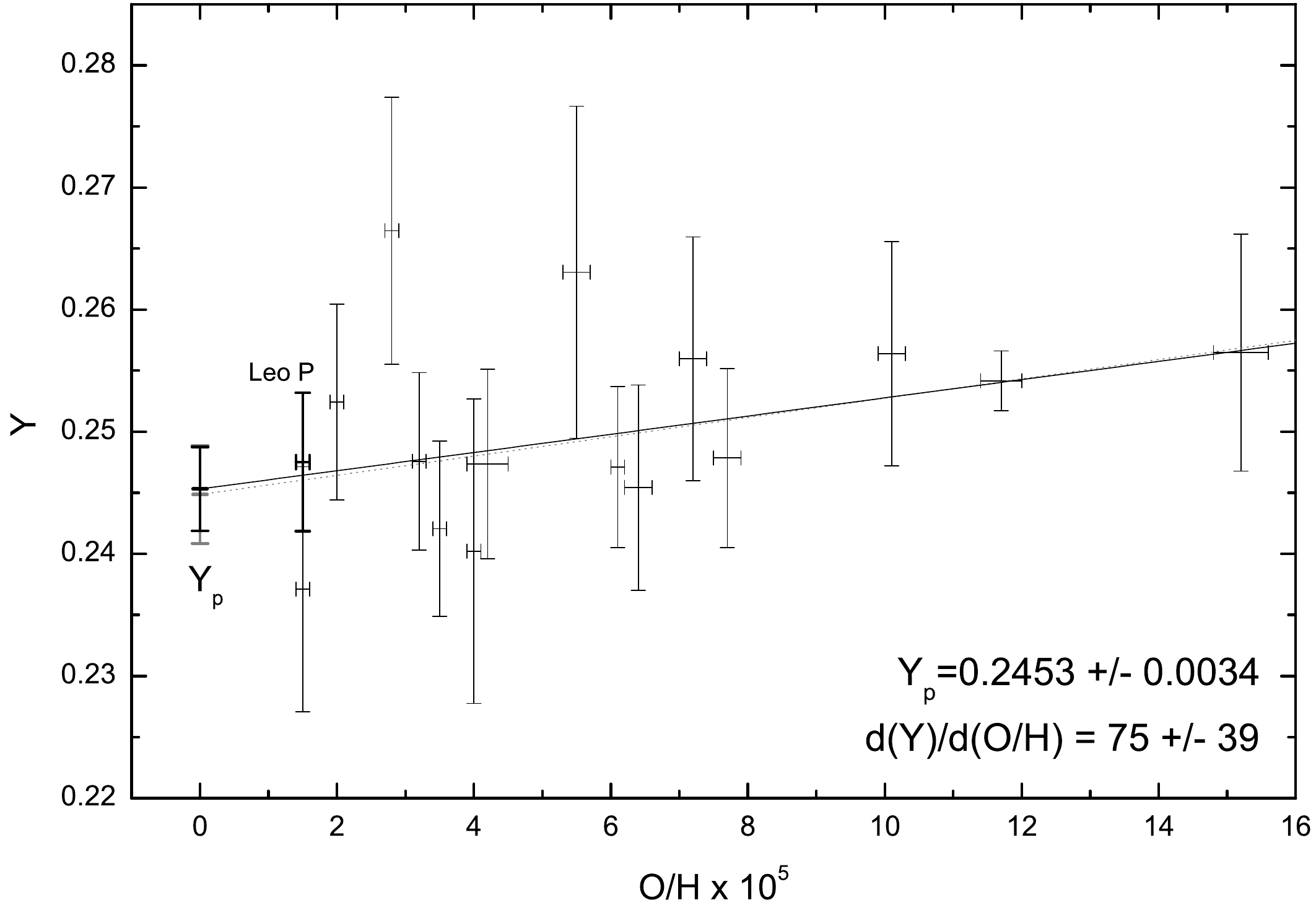}
\caption{
Helium abundance (mass fraction) versus oxygen to hydrogen ratio regression calculating the primordial helium abundance.  
The previous result is lighter grey, with the new result, based on including Leo~P, shown with a solid fit line and a bolded intercept. Leo~P is also shown as bold.  }
\label{Y-OH}
\end{center}
\end{figure}

Most recently, it was shown that adding an additional 2 He lines, and nine H lines to fit nine input parameters can greatly improve the the accuracy
of the fit \cite{LeoP}. An example of the potential improvement was demonstrated on data from Leo P.
Table~\ref{table:LeoP} shows the improvement made in going from nine emission line ratios to fit eight parameters made in 2013 \cite{skillman2013}, to 21 emission line ratios to fit nine parameters. All of the parameters are found with smaller uncertainties (sometimes significantly smaller) and a factor of 3 improvement in the helium mass fraction uncertainty. The primordial \he4 abundance is extracted from a linear fit to the data with respect to the oxygen abundance and is shown in Fig.~\ref{Y-OH} \cite{LeoP}. The resulting primordial abundance is found to be
$Y_P = 0.2453 \pm 0.0034$, a 15\% improvement solely due to Leo P. This is to be compared with the uncertainty of 0.0002 in the BBN calculation.

\begin{table}[ht!]
\caption{Physical conditions, He$^+$/H$^+$ abundance solution, and regression values of Leo~P}
\label{table:LeoP}
\centering
\begin{tabular}{|lcc|}
\hline\hline
                        & Skillman et al. (2013) \cite{skillman2013}              &  Aver et al. (2021) \cite{LeoP}                  \\
\hline
Emission lines          &  9                                &  21                           \\
Free Parameters         &  8                                &  9                            \\
\hline
He$^+$/H$^+$		    &  0.0837$^{+0.0084}_{-0.0062}$     &  0.0823$^{+0.0025}_{-0.0018}$ \\
n$_e$ [cm$^{-3}$]	    &  1$^{+206}_{-1}$                  &  39$^{+12}_{-12}$             \\
a$_{He}$ [\AA]		    &  0.50$^{+0.42}_{-0.42}$           &  0.42$^{+0.11}_{-0.15}$       \\
$\tau$				    &  0.00$^{+0.66}_{-0.00}$           &  0.00$^{+0.13}_{-0.00}$       \\
T$_e$ [K]			    &  17,060 $^{+1900}_{-2900}$        &  17,400 $^{+1200}_{-1400}$    \\
C(H$\beta$)			    &  0.10$^{+0.03}_{-0.07}$           &  0.10$^{+0.02}_{-0.02}$       \\
a$_H$ [\AA]			    &  0.94$^{+1.44}_{-0.94}$           &  0.51$^{+0.17}_{-0.18}$       \\
a$_P$ [\AA]             &  -                                &  0.00$^{+0.52}_{-0.00}$       \\
$\xi$ $\times$ 10$^4$   &  0$^{+156}_{-0}$                  &  0$^{+7}_{-0}$                \\
$\chi^2$			    &  3.3                              &  15.3                         \\
p-value                 &  7\%                              &  23\%                         \\
\hline
O/H $\times$ 10$^5$	    &  1.5 $\pm$ 0.1                    &  1.5 $\pm$ 0.1                \\
Y				        &  0.2509 $\pm$ 0.0184              &  0.2475 $\pm$ 0.0057          \\
\hline
\end{tabular}
\end{table}

\section{The impact of New Cross Section Measurements} 
\begin{figure}[!htb]
\begin{center}
\includegraphics[width=0.35\textwidth]{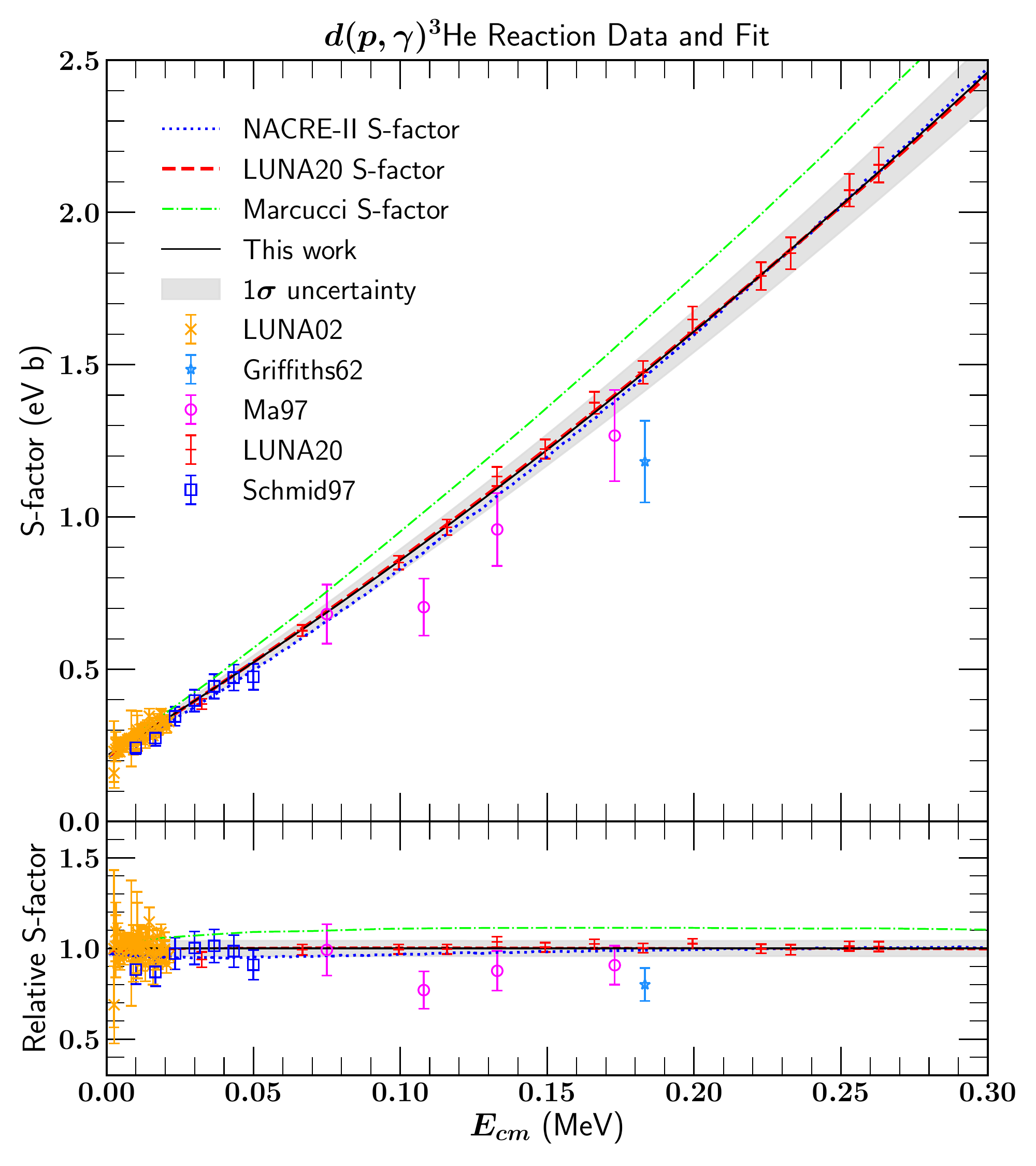}
\caption{
The astrophysical S-factor for $d(p,\gamma)$\he3 showing 1) the NACRE-II  $S$-factor used in FOYY (blue dotted); 2) the theoretical $S$-factor (green dot-dashed); 3) the LUNA global average  (red dashed); and 4) our new world average rate (black solid). The shading corresponds to the 68\% uncertainty. 
    \label{fig:sfac}
}
\end{center}
\end{figure}

In contrast to \he4, the observed uncertainty in D/H is $3 \times 10^{-7}$ compared with the theoretical uncertainty of $1.1 \times 10^{-6}$,
due to uncertainties in the experimental cross sections.  Recently, the \dpg cross section was remeasured by the LUNA collaboration \cite{LUNA}
in the BBN energy range with significantly higher accuracy than previous measurements. This data is shown in Fig.~\ref{fig:sfac}
along with previous data as labeled. Also shown are several fits: one based on NACRE-II data~\cite{nacreII}, a theory-based cross section \cite{marc}, the LUNA collaboration fit \cite{LUNA} and our fit including previous data~\cite{YOF}. The latter two are very similar, as the fit is driven by LUNA data.

While the new \dpg cross section affects slightly the abundances of \he4 and \li7, the dominant effect is on D/H, as might be expected. 
The effect of the \dpg cross section on the CMB-BBN likelihood functions is shown in the right set of figures in Fig.~\ref{fig:2x2abs_2d}.
In the upper left is the previous result (as in the upper right of the left panel) \cite{FOYY}. The theory cross section,
has a significantly larger rate and therefore predicts substantially less D/H. The two bottom figures are based on the LUNA fit and 
the combined fit \cite{YOF}.  The mean value of D/H ($\times 10^5$) for each case is $2.57 \pm .13$, $2.42 \pm .10$, $2.50 \pm .11$
and $2.51 \pm .11$.  The new cross section shifts slightly the best value for $\eta_{10}$ to $6.123 \pm 0.039$. The residual error is due to the remaining uncertainty in $d(d,n) {}^3$He and $d(d,p) {}^3$H. For related studies of this new cross section, see Ref. \cite{Pitrou2020}. 

When $N_\nu$ is not held fixed at the Standard Model value of 3, the likelihood analysis leads to a mean value of $N_\nu = 2.880 \pm 0.144$
implying a 95\% upper limit of $N_\nu < 3.16$ and provides strong constraints on physics beyond the Standard Model.

\section*{Acknowledgments}
I thank E. Aver, B. Fields, E. Skillman, and T.-H. Yeh for fruitful collaborations that lead to the results presented here. 
This work was supported in part by DOE grant DE-SC0011842.


\section*{References}

\begin{thebibliography}{99}

\bibitem{bbn} K.~A.~Olive, G.~Steigman and T.~P.~Walker,
  Phys.\ Rept.\  {\bf 333}, 389 (2000)
  [astro-ph/9905320].

 \bibitem{cfo1}
R.~H.~Cyburt, B.~D.~Fields and K.~A.~Olive,
{\it New Astron.\  } {\bf 6} (1996) 215
[arXiv:astro-ph/0102179].

 \bibitem{coc}
  E.~Vangioni-Flam, A.~Coc and M.~Casse,
  Astron.\ Astrophys.\  {\bf 360}, 15 (2000)
  [astro-ph/0002248];
A.~Coc, E.~Vangioni-Flam, P.~Descouvemont, A.~Adahchour and C.~Angulo,
{\it Ap.\ J.\  } {\bf 600} (2004) 544
[arXiv:astro-ph/0309480].

\bibitem{cyburt}
R.~H.~Cyburt,
Phys.\ Rev.\ D {\bf 70} (2004) 023505
[arXiv:astro-ph/0401091].

\bibitem{iocco}
  F.~Iocco, G.~Mangano, G.~Miele, O.~Pisanti and P.~D.~Serpico,
  Phys.\ Rept.\  {\bf 472}, 1 (2009)
  [arXiv:0809.0631 [astro-ph]].
  
  \bibitem{CFOY}
R.~H.~Cyburt, B.~D.~Fields, K.~A.~Olive and T.-H. Yeh,
  Rev.\ Mod.\ Phys.\  {\bf 88}, 015004 (2016)
  [arXiv:1505.01076 [astro-ph.CO]].

  
   \bibitem{coc18}
   C.~Pitrou, A.~Coc, J.~P.~Uzan and E.~Vangioni,
  Phys.\ Rept.\  {\bf 754}, 1 (2018)
  [arXiv:1801.08023 [astro-ph.CO]].
  
\bibitem{FOYY}
B.~D.~Fields, K.~A.~Olive, T.~H.~Yeh and C.~Young,
JCAP \textbf{03}, 010 (2020)
[erratum: JCAP \textbf{11}, E02 (2020)]
[arXiv:1912.01132 [astro-ph.CO]].

 \bibitem{kk}
  P.~J.~Kernan and L.~M.~Krauss,
  Phys.\ Rev.\ Lett.\  {\bf 72}, 3309 (1994)
  [astro-ph/9402010];
  L.~M.~Krauss and P.~J.~Kernan,
  Phys.\ Lett.\ B {\bf 347}, 347 (1995)
  [astro-ph/9408023];
  C.~J.~Copi, D.~N.~Schramm and M.~S.~Turner,
  Phys.\ Rev.\ D {\bf 55} (1997) 3389
  [astro-ph/9606059];
  K.~A.~Olive and D.~Thomas,
  Astropart.\ Phys.\  {\bf 7} (1997) 27
  [hep-ph/9610319];
  K.~A.~Olive and D.~Thomas,
  Astropart.\ Phys.\  {\bf 11}, 403 (1999)
  [hep-ph/9811444].
    
    \bibitem{lisi}
   E.~Lisi, S.~Sarkar and F.~L.~Villante,
  Phys.\ Rev.\ D {\bf 59}, 123520 (1999)
  [hep-ph/9901404];
   S.~Sarkar,
  Rept.\ Prog.\ Phys.\  {\bf 59}, 1493 (1996)
  [hep-ph/9602260].

  
    \bibitem{cfos}
    R.~H.~Cyburt, B.~D.~Fields, K.~A.~Olive and E.~Skillman,
  Astropart.\ Phys.\  {\bf 23}, 313 (2005)
  [astro-ph/0408033].
  
  \bibitem{ms}
  G.~Mangano and P.~D.~Serpico,
  Phys.\ Lett.\ B {\bf 701}, 296 (2011)
  [arXiv:1103.1261 [astro-ph.CO]].



\bibitem{wmap1}
  D.~N.~Spergel {\it et al.}  [WMAP Collaboration],
  Astrophys.\ J.\ Suppl.\  {\bf 148}, 175 (2003)
  [astro-ph/0302209].
  
  \bibitem{Planck2018}
  N.~Aghanim \textit{et al.} [Planck],
Astron. Astrophys. \textbf{641}, A6 (2020)
[arXiv:1807.06209 [astro-ph.CO]].
  
  \bibitem{cfo2}
R.~H.~Cyburt, B.~D.~Fields and K.~A.~Olive,
Astropart.\ Phys.\  {\bf 17} (2002) 87
[arXiv:astro-ph/0105397].

 \bibitem{kr}
L.~M.~Krauss and P.~Romanelli,
  Astrophys.\ J.\  {\bf 358}, 47 (1990);
  M.~S.~Smith, L.~H.~Kawano and R.~A.~Malaney,
  Astrophys.\ J.\ Suppl.\  {\bf 85}, 219 (1993).
  
   
  \bibitem{fo}
  B.~D.~Fields and K.~A.~Olive,
  Phys.\ Lett.\ B {\bf 368}, 103 (1996)
  [hep-ph/9508344];
 B.~D.~Fields, K.~Kainulainen, K.~A.~Olive and D.~Thomas,
  New Astron.\  {\bf 1}, 77 (1996)
  [astro-ph/9603009].


\bibitem{cfo3}
 R.~H.~Cyburt, B.~D.~Fields and K.~A.~Olive,
  Phys.\ Lett.\ B {\bf 567}, 227 (2003)
  [astro-ph/0302431].
  
     \bibitem{cfo5}
   R.~H.~Cyburt, B.~D.~Fields and K.~A.~Olive,
  JCAP {\bf 0811}, 012 (2008)
  [arXiv:0808.2818 [astro-ph]].
  
  \bibitem{its07} 
  Y.~I.~Izotov, T.~X.~Thuan and G.~Stasi\'nska,
  Astrophys.\ J.\  {\bf 662}, 15 (2007)
  [arXiv:astro-ph/0702072].

\bibitem{AOS2}
  E.~Aver, K.~A.~Olive and E.~D.~Skillman,
  JCAP {\bf 1103}, 043 (2011)
  [arXiv:1012.2385 [astro-ph.CO]] (AOS2).

\bibitem{AOS3}
  E.~Aver, K.~A.~Olive and E.~D.~Skillman,
  JCAP {\bf 1204}, 004 (2012)
  [arXiv:1112.3713 [astro-ph.CO]] (AOS3).
  
  \bibitem{itg14} 
  Y.~I.~Izotov, T.~X.~Thuan and N.~G.~Guseva,
  Mon.\ Not.\ Roy.\ Astron.\ Soc.\  {\bf 445}, 778 (2014)
  [arXiv:1408.6953 [astro-ph.CO]] (ITG14).

\bibitem{AOS4}
  E.~Aver, K.~A.~Olive and E.~D.~Skillman,
  JCAP \textbf{07}, 011 (2015)
  [arXiv:1503.08146 [astro-ph.CO]] (AOS4).

\bibitem{LeoP}
E.~Aver, D.~A.~Berg, K.~A.~Olive, R.~W.~Pogge, J.~J.~Salzer and E.~D.~Skillman,
JCAP \textbf{03}, 027 (2021)
[arXiv:2010.04180 [astro-ph.CO]].

\bibitem{skillman2013} 
  E.~D.~Skillman, J.~J.~Salzer, D.~A.~Berg, R.~W.~Pogge, N.~C.~Haurberg, J.~M.~Cannon, E.~Aver, K.~A.~Olive, R.~Giovanelli, M.~P.~Haynes, E.~A.~K.~Adams, K.~B.~W.~McQuinn and K.~L.~Rhode,
  Astron. J. \textbf{146}, 3 (2013)
  [arXiv:1305.0277 [astro-ph.CO]].
  
  \bibitem{LUNA}
V.~Mossa, K.~St\"ockel, F.~Cavanna, F.~Ferraro, M.~Aliotta, F.~Barile, D.~Bemmerer, A.~Best, A.~Boeltzig and C.~Broggini, \textit{et al.}
Nature \textbf{587}, 210 (2020).

 \bibitem{nacreII}
  Y.~Xu, K.~Takahashi, S.~Goriely, M.~Arnould, M.~Ohta and H.~Utsunomiya,
  Nucl.\ Phys.\ A {\bf 918}, 61 (2013)
  [arXiv:1310.7099 [nucl-th]].
  
  \bibitem{marc}
 L.~E.~Marcucci, G.~Mangano, A.~Kievsky and M.~Viviani,
  Phys.\ Rev.\ Lett.\  {\bf 116}, no. 10, 102501 (2016)
  Erratum: [Phys.\ Rev.\ Lett.\  {\bf 117}, no. 4, 049901 (2016)]
  [arXiv:1510.07877 [nucl-th]].
  
\bibitem{YOF}
T.~H.~Yeh, K.~A.~Olive and B.~D.~Fields,
JCAP \textbf{03}, 046 (2021)
[arXiv:2011.13874 [astro-ph.CO]].

\bibitem{Pitrou2020}
C.~Pitrou, A.~Coc, J.~P.~Uzan and E.~Vangioni,
Mon. Not. Roy. Astron. Soc. \textbf{502}, no.2, 2474-2481 (2021)
[arXiv:2011.11320 [astro-ph.CO]];
O.~Pisanti, G.~Mangano, G.~Miele and P.~Mazzella,
JCAP \textbf{04}, 020 (2021)
[arXiv:2011.11537 [astro-ph.CO]].


\end{thebibliography}
\end{document}